\documentclass[9pt,conference]{IEEEtran}
\IEEEoverridecommandlockouts
\usepackage{cite}
\usepackage{amsmath,amssymb,amsfonts}
\usepackage{algorithmic}
\usepackage{graphicx}
\usepackage{steinmetz}
\usepackage{tabularx}
\usepackage{textcomp}
\usepackage{float}
\usepackage{booktabs}
\usepackage{array}
\usepackage{mathtools}
\usepackage{threeparttable}
\usepackage{balance}
\usepackage{url}
\usepackage{hyperref}
\usepackage[table]{xcolor}
\usepackage{enumitem}
\newcolumntype{M}[1]{>{\centering\arraybackslash}m{#1}}

\newcolumntype{C}{>{\centering\arraybackslash}p{5cm}}
\def\BibTeX{{\rm B\kern-.05em{\sc i\kern-.025em b}\kern-.08em
    T\kern-.1667em\lower.7ex\hbox{E}\kern-.125emX}}
\usepackage{pifont}
\newcommand{\cmark}{\ding{51}}
\newcommand{\xmark}{\ding{55}}
\begin{document}

\title{\textbf{AaLLM}: An End-to-End \textbf{A}n\textbf{a}log Circuit Design Framework from Topology Generation to Sizing Using \textbf{L}arge \textbf{L}anguage \textbf{M}odels}
\author{\IEEEauthorblockN{Mohammed~Ayman~Habib, Rylan~Hart, Morteza~Fayazi}
Email: m.habib@utah.edu, rylan.hart@utah.edu, m.fayazi@utah.edu
\IEEEauthorblockA{\textit{University of Utah, Salt Lake City, Utah, USA}}}
\maketitle
\begin{abstract}
Analog circuit design is a time-consuming, iterative process in a nonlinear and high-dimensional design space that relies heavily on expert intuition. Among recent developments, Large Language Models (LLMs) have introduced a promising approach by bringing natural language reasoning to circuit design tasks. The majority of conventional LLM-based approaches provide fragmented solutions that focus either only on sizing or topology generation. These methods require adding specific technical knowledge manually, which is inefficient and prone to hallucinations during circuit sizing. Moreover, the inherent trade-off in meeting different specs makes current approaches iterative and tedious. Another shortcoming is the inability to create innovative topologies, which may lead to sub-optimal designs due to reliance on conventional topologies. In this paper, we present AaLLM, an open-source end-to-end multi-agent LLM workflow that takes user specs as input and outputs the appropriate netlist, encompassing both topology generation and circuit sizing. AaLLM automates the creation of a relevant knowledge base from research papers and textbooks to combat tedious manual data collection. A Retrieval Augmented Generation (RAG) model is implemented to emulate circuit design expertise using this knowledge base. Moreover, AaLLM uses a novel tri-agent feedback system comprising a Designer that determines circuit component values, a Critic that scrutinizes these values, and an Evaluator that minimizes circuit sizing iterations by arbitrating between the other two agents. Additionally, AaLLM generates novel, valid topologies by leveraging a fine-tuned Sequence-to-Sequence (Seq2Seq) model that learns the effect of each connection in conventional topologies and recombines them to form new circuits. AaLLM-generated novel topologies achieve a figure of merit (FoM) comparable to that of known topologies, and up to 3x higher for certain circuits. Testing on several circuit topologies, our results show a 3x - 4.5x decrease in the number of SPICE calls at inference when compared to state-of-the-art multi-agent LLM pipelines. The results also show a 40x decrease in wall-clock time compared to existing approaches.
\end{abstract}


\section{Introduction}
The increasing complexity of modern analog circuits, combined with strict performance requirements and process variations, has made manual circuit design impractical and time-consuming~\cite{fayazi_2021}. Therefore, the need for automation in analog circuit design cannot be overstated. The design process is subdivided into topology selection and sizing of circuit elements to achieve the target specs. 

Machine Learning (ML) methods have shown significant promise in analog circuit sizing. Techniques such as Neural Network (NN)-based surrogate models, Graph Neural Networks (GNN), and Reinforcement Learning (RL) have been applied for this purpose with notable improvements in speed and accuracy~\cite{Ren_2020, Settaluri_2020, brempong2026oracle, fayazi2022fascinet}. However, most of them learn direct mappings from performance specs (\textit{e.g.}, gain, power, bandwidth) to circuit design parameters (\textit{e.g.}, component values, transistor sizes) without capturing the underlying design reasoning. In other words, these models operate as black boxes, offering no insight into why a particular set of design parameters was chosen. This lack of reasoning makes it difficult to diagnose failures or adapt the design strategy when specs are not met. Beyond ML, optimization-based approaches have also been widely used for circuit sizing~\cite{Lyu_2018}. While these methods can achieve high accuracy, they rely on extensive SPICE simulation iterations, making them computationally expensive and time-consuming. 

Many existing approaches focus solely on sizing while leaving the topology generation problem unaddressed. Topology generation is challenging as it requires understanding how different circuit structures affect performance trade-offs. Existing methods for topology generation either select from a small predefined library or construct circuits from basic building blocks without leveraging any understanding of circuit physics~\cite{fayazi2023angel}. The former is constrained by the limited number of topologies, preventing exploration of designs beyond what has been cataloged. On the other hand, the latter produces a large number of candidate topologies, but it lacks an efficient way to evaluate which ones are electrically functional, resulting in long synthesis times. These limitations highlight the need for a unified approach that handles both topology generation and circuit sizing within a single framework.

Large Language Models (LLMs) offer a viable solution to these challenges~\cite{abbineni2026muallm, aldowaish2026sina} by reasoning about analog circuit behavior in natural language, much as an experienced designer would. However, most of these approaches treat sizing as a single-shot generation task without simulation feedback, determining circuit parameter values based on pervasive pattern recognition rather than circuit reasoning. This often leads to hallucinated sizing decisions that fail to meet specs or violate device physics. Furthermore, existing LLM-based approaches lack the ability to diagnose why a design fails; they can identify that a spec is unmet, but cannot trace the root cause to a specific circuit component. Without iterative refinement and diagnostic reasoning, these methods often improve one spec at the expense of another. 

To combat the aforementioned challenges, we propose AaLLM, an open-source\footnote{\url{https://anonymous.4open.science/r/AaLLM/README.md}} end-to-end analog circuit design framework that spans from topology generation to sizing, leveraging LLMs. The goals of AaLLM are threefold: (a) automate the retrieval of relevant design knowledge from analog circuit literature to inform decision-making at every stage of the design process; (b) generate novel, valid analog circuit topologies that go beyond textbook designs; and (c) provide a unified framework that handles both topology generation and circuit sizing within a single automated pipeline. 

A key component supporting these goals is the Retrieval-Augmented Generation (RAG) module, which provides the framework with access to relevant design knowledge~\cite{lewis2020rag}. This module is used to evaluate and select the most suitable topology for the given specs. The retrieval module searches a curated database of analog circuit design literature to provide the agents with relevant design knowledge during both topology selection and sizing. This allows the agents to ground their decisions in established circuit theory rather than relying solely on patterns learned during training. As a result, AaLLM handles unfamiliar spec combinations by drawing on relevant prior work, rather than being limited to what it has seen during fine-tuning. Furthermore, the retrieval module autonomously expands its knowledge base by fetching and processing new research papers from academic repositories, ensuring the agents have access to up-to-date design information. This dynamic update capability eliminates the need for manual curation, keeping the system up to date with the latest research in the field.

To support novel topology generation, AaLLM employs a fine-tuned~\cite{devlin2019bert} generative model that operates over the full combinatorial space of component connections. This allows the framework to produce both conventional and novel topologies that may not appear in existing literature. The generated candidates are then evaluated for feasibility through the RAG module before being passed to the sizing stage, ensuring that only viable topologies proceed to optimization.

For circuit sizing, AaLLM employs a tri-agent architecture comprising a Designer, a Critic, and an Evaluator, which work together in an iterative loop to refine the circuit parameters. The loop relies on SPICE simulations, providing accurate circuit performance specs for the agents to reason over. At each iteration, the Designer determines circuit component values, the Critic scrutinizes these values, and the Evaluator minimizes circuit sizing iterations by arbitrating between the other two agents. The optimization follows a curriculum-based approach, where design objectives are addressed in a deliberate sequence, allowing each spec to be met without disrupting previously satisfied ones.

AaLLM pairs two distinct types of language models: one fine-tuned specifically for topology generation and another that collaboratively reasons through a multi-agent system to perform circuit sizing. This combination allows each model to be applied where it is most effective. The fine-tuned model captures topology-level patterns from circuit data, while the multi-agent system brings flexible circuit reasoning to the sizing process. To further enhance the agents' reasoning capabilities, AaLLM incorporates RAG, in which relevant design knowledge from analog circuit textbooks and technical papers is retrieved and used as context during both topology generation and sizing. This enables the agents to ground their decisions in established circuit theory, allowing them to make informed design choices even for unfamiliar topologies or spec combinations. For topology generation, AaLLM employs a Sequence-to-Sequence (Seq2Seq) model that translates a set of desired specs into a valid circuit netlist. By treating topology generation as a translation task, the model learns structural patterns across a wide range of circuit designs, enabling it to produce both conventional and novel topologies. Seq2Seq models are well-suited for this task as they naturally capture the dependencies between circuit elements. Moreover, fine-tuning on circuit-specific data enables the model to learn the structural constraints and connectivity rules that govern valid analog topologies, yielding more reliable outputs than prompting a general-purpose model. 

To validate our proposed method, AaLLM is tested by generating and sizing op-amp and filter topologies. AaLLM-generated novel topologies achieve a figure of merit (FoM) comparable to that of known topologies, and up to \textbf{3x} higher for certain circuits. Testing on several circuit topologies, our results show a \textbf{3x - 4.5x} decrease in the number of SPICE calls at inference when compared to state-of-the-art multi-agent LLM pipelines. The results also show a \textbf{40x} decrease in wall-clock time compared to State-Of-The-Art (SOTA) approaches. Additionally, the tri-agent sizing loop meets the target specs in \textbf{91.6\%} of cases across a test bench spanning a wide range of design specs.

The main contributions of this paper can be summarized as follows:
\begin{itemize}[topsep=0pt, itemsep=2pt, leftmargin=*]
 \item Retrieval-augmented generation applied at both topology selection and circuit sizing, grounding design decisions in established circuit theory at every stage of the pipeline.
 \item Generation of novel circuit topologies that go beyond known designs, with candidates validated through both theoretical evaluation and SPICE simulation.
 \item A tri-agent sizing architecture that separates diagnosis, strategy, and parameter adjustment into distinct roles, enabling the system to identify why a spec is not met and correct it systematically.
 \item An open-source, fully autonomous end-to-end framework that connects topology generation, selection, and circuit sizing into a unified pipeline where each stage informs the next.
\end{itemize}

\section{Background \& Related Work}

\subsection{Problem Formulation}
The goal of analog circuit design in AaLLM is to jointly determine the circuit topology $T$ and the circuit design parameter vector $x$ (\textit{e.g.}, DC bias voltages and size of transistors) such that
\begin{equation}
\begin{aligned}
\text{find} \quad & T \in \mathcal{P}, \; x \in \mathbb{R}^n \\
\text{such that:} \quad & c_1(T, x) < 0, \ldots, c_N(T, x) < 0,
\end{aligned}
\label{eq:aallm_opt}
\end{equation}
where $T$ is a circuit topology drawn from the space of valid topologies $\mathcal{P}$, $x$ is the parameter vector for that topology, and $c_1, \ldots, c_N$ are user-supplied constraints on the circuit's performance metrics.

LLMs are deep neural networks based on the transformer architecture. They process sequential data through a self-attention mechanism that captures long-range dependencies between tokens in an input sequence~\cite{vaswani2017attention}. Models such as GPT-4, LLaMA, and DeepSeek are pre-trained on massive text corpora using next-token prediction, enabling them to learn rich representations of language, logic, and domain-specific knowledge. After pre-training, LLMs can be adapted to specialized tasks through supervised fine-tuning on curated datasets, as demonstrated by AnalogSeeker~\cite{chen2025analogseeker} and LaMagic~\cite{chang2024lamagic}. Moreover, as employed by AMSnet-KG~\cite{shi2025amsnet} and AmpAgent~\cite{liu2024ampagent}, a RAG module is leveraged to incorporate external knowledge sources such as research papers. A key capability of LLMs is in-context learning, where the model generalizes to new tasks from a few examples provided in the prompt without updating its parameters, as leveraged by AnalogCoder~\cite{lai2025analogcoder} in its training-free design flow. When augmented with an external feedback tool (\textit{e.g.} SPICE), code execution platform (\textit{e.g.} PySPICE), and multi-agent orchestration, LLMs become agentic systems capable of iterative reasoning, planning, and interaction. These properties have motivated LLM adoption in electronic design automation, where they serve as reasoning engines, code generators, or optimization agents for traditionally manual design workflows.

\subsection{Topology Generation}
\label{sec:two-port_analysis}
AnalogGenie~\cite{gao2025analoggenie} employs a GPT-based generative model that discovers diverse topologies using a sequential graph representation based on Eulerian circuits. AnalogXpert~\cite{zhang2025analogxpert} leverages reusable sub-circuit libraries and experience-based proofreading strategies to decompose topology synthesis into sub-circuit selection and connection. LaMagic~\cite{chang2024lamagic} applies supervised fine-tuning with novel circuit formulations, including adjacency-matrix-based and float-input representations, to generate optimized power-converter topologies. AutoCircuitRL~\cite{vijayaraghavan2025autocircuit} combines instruction tuning with iterative feedback-driven refinement to generate netlists.

\subsection{Circuit Sizing}
AnaFlow~\cite{ahmadzadeh2025anaflow} proposes a multi-phase agentic workflow with specialized agents for DC operating point analysis, reasoning-driven optimization, and simulator-equipped refinement. AmpAgent~\cite{liu2024ampagent} deploys a multi-agent system that extracts sizing formulas from literature using RAG to size multi-stage amplifiers. AutoSizer~\cite{yu2026autosizer} introduces a two-loop multi-agent framework where an inner loop performs parameter optimization and an outer loop analyzes optimization dynamics. EasySize~\cite{wu2025easysize} combines LLM-guided heuristic search with differential evolution and particle swarm optimization using an ease-of-attainability spec. LEDRO~\cite{kochar2025ledro} uses LLMs iteratively to refine the design search space through calibration point synthesis, coupling LLM exploration with Bayesian optimization. Ghosh \textit{et~al.}~\cite{ghosh2025accelerating} employ transformer models with precomputed lookup tables to map amplifier specs to transistor parameters with minimal SPICE simulations.

\subsection{Topology Generation + Circuit Sizing}
A growing body of work addresses both topology generation and circuit sizing within unified frameworks. AnalogCoder~\cite{lai2025analogcoder} leveraged a training-free approach where an LLM agent generates Python code using feedback-enhanced verification to iteratively correct the designs. AnalogCoder-Pro~\cite{lai2026analogcoder} extends this with multimodal inputs, including waveform analysis. Artisan~\cite{chen2024artisan} employs a domain-specific LLM with bidirectional circuit representations combining Tree-of-Thoughts and Chain-of-Thoughts (CoT) reasoning for topology and parameter tuning. Atelier~\cite{shen2025atelier} deploys five specialized LLM agents within a Graph-of-Thoughts architecture to decompose hierarchical design tasks. LADAC~\cite{liu2024ladac} uses an LLM agent guided by a local knowledge library for iterative topology and sizing co-design. MenTeR~\cite{chen2025menter} presents a fully automated multi-agent workflow decomposing user prompts into actual specs. Moreover, it generates proper testbenches with CoT reasoning.

\section{Proposed Approach}
\begin{figure*}[t]
\centering
\includegraphics[width=\textwidth]{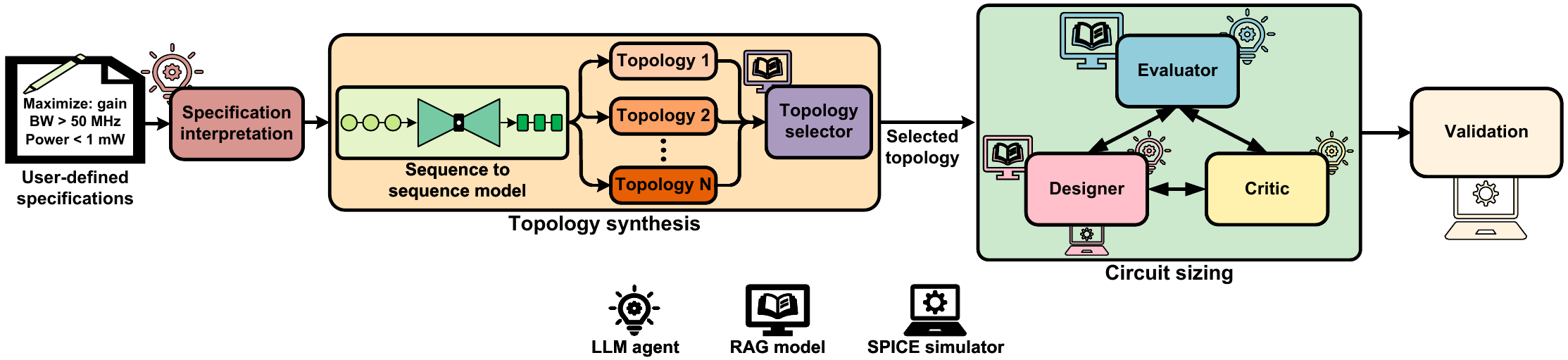}
\caption{High-level platform architecture of AaLLM.}
\label{fig:framework}
\end{figure*}

\subsection{Overview}\label{sec:overview}
As depicted in Fig.~\ref{fig:framework}, AaLLM decomposes the analog design problem into two complementary stages handled by different types of LLMs: a fine-tuned generative model for topology generation, and a multi-agent reasoning system for circuit sizing. Both stages are supported by the RAG module that draws on established circuit theory from an analog design knowledge base. 

\subsection{Circuit Representation}\label{sec:representation}
AaLLM represents every circuit as a \emph{bipartite component-node matrix}~\cite{chang2024lamagic} rather than a free-form SPICE netlist~\cite{lai2025analogcoder} or a directed graph~\cite{dong2023cktgnn}. Every connection in a circuit links a component terminal to a circuit node. This natural component-to-node relationship forms a bipartite graph $G = (\mathcal{C} \cup \mathcal{N}, E)$, where $\mathcal{C}$ is the set of components, $\mathcal{N}$ is the set of nodes, and an edge $e \in E$ exists whenever a component terminal connects to a node. The graph can be encoded as a bipartite matrix $\mathbf{B} \in \mathcal{V}^{|\mathcal{C}| \times |\mathcal{N}|}$, where each entry $B_{ij}$ specifies which terminal of component $c_i$ connects to node $n_j$. This component-node matrix serves two purposes: it decouples topology from sizing, and it provides a fixed-length token structure that the transformer decoder can generate under position-aware constraints. Moreover, AaLLM guarantees syntactically valid circuit generation, because every token sits in a known position within a finite-length token structure. Since every sequence length is fixed, the \textit{LogitsProcessor}~\cite{wolf2020transformers}, a feature of the LLM responsible for token generation, knows exactly what position it is at. The serialized bipartite matrix sequence consists of four sections separated by \texttt{<sep>} delimiters. The first is a component-type vector $\boldsymbol{\tau}$ listing the type of each component \textit{e.g.} \texttt{NMOS}, \texttt{CAP}, \texttt{<none>}, etc. The second and third are fixed lists of component names (\textit{e.g.} \texttt{M1}, \texttt{C1}, etc.) and net names. The fourth is a flattened connection matrix, in which each component row is prefixed by its name and followed by one connection token per net. 

\subsection{Specification Resolution}
\label{sec:spec_resolution}
The inputs to AaLLM are the performance requirements that a circuit must meet. These are typically specified by the user as a set of bounds on specs such as gain, bandwidth, and power. Such requirements are naturally expressed as inequality constraints, for example a minimum gain or a maximum power budget. However, AaLLM requires a single concrete spec point to generate against. It is conditioned on fixed scalar values during fine-tuning and cannot directly interpret an inequality. 

The specification resolver bridges this gap. It projects the user's constraint set onto the region of spec space that the topology generator has seen during fine-tuning, and returns a single target point within that region. The resolver operates independently on each performance spec $k$. For each spec, it first identifies the range of values seen during fine-tuning, $[\underline{t}_k, \bar{t}_k]$. It then intersects this range with the user's specified constraint, $[\underline{c}_k, \bar{c}_k]$, to obtain the effective feasible interval:
\begin{equation}\label{eq:feasible_interval}
    [\ell_k,\,u_k] = \bigl[\max(\underline{c}_k,\underline{t}_k),\;
        \min(\bar{c}_k,\bar{t}_k)\bigr].
\end{equation}

The target value for the spec is placed at the midpoint of this interval:
\begin{equation}\label{eq:midpoint_target}
    s^{*}_k = \frac{\ell_k + u_k}{2}.
\end{equation}
The midpoint is chosen because it maximizes the distance from both boundaries of the training range. This minimizes the risk of querying AaLLM near the edge of its training range where performance degrades. Moreover, this ensures AaLLM is queried at a point which falls in both the user's constraints and the training range. Unspecified specs default to the training-set mean, serving as a neutral anchor when no user preference is given. If a constraint falls outside the training range, the resolver clamps $s^{*}_k$ to the nearest boundary and emits a warning.

\subsection{Topology Generation}\label{sec:topology_generation}
The topology generator maps the resolved target $\mathbf{s}^{*}$ to a set of candidate circuit structures. Unlike parameter sizing, this is a fundamentally discrete task: the model must decide which circuit components to include and how to connect them together. We therefore formulate it as a conditional Sequence-to-Sequence (Seq2Seq) problem, learning the mapping from performance specs to bipartite matrices.

The target spec ($\mathbf{s}^{*}$) is first converted into a natural-language prefix. This prefix comprises each spec as a text label and its corresponding value. This prefix is then concatenated with a template of the bipartite matrix. In the template, the component-type and connection positions are replaced by placeholder tokens that mark where the model must predict a value. The prefix tells the model what to design, and the template provides the format in which the answer should be written. The combined sequence is fed to a fine-tuned FLAN-T5 encoder-decoder~\cite{chang2024lamagic}. The decoder then fills in the placeholders auto-regressively to produce a complete bipartite matrix conditioned on the target spec.

We adopt FLAN-T5 for three reasons. First, its encoder-decoder split naturally separates spec conditioning from topology generation. Second, its instruction-tuned pre-training provides a strong initialization for constrained sequence tasks. Third, FLAN-T5 was pre-trained using a fill-in-the-blank style objective known as span corruption, in which a model is shown text sequences with random chunks of words hidden and has to predict the missing chunks from the surrounding context. This is precisely aligned with the structure of our task: the bipartite matrix template contains blank positions that the model must fill in based on the surrounding structure and the target spec. 

Training follows a three-stage curriculum that gradually exposes the model to harder versions of the task. In the first stage, the component-type vector $\boldsymbol{\tau}$ is revealed and only the connection entries are masked. The model therefore learns connectivity patterns conditioned on a known set of components. The training data is further augmented by randomly reordering the component labels in each circuit while keeping the underlying connections unchanged. This teaches the model that a circuit's function depends only on how its components are connected, not on the order in which they happen to be listed.

The second stage drops the permutation augmentation but keeps the same masking. This allows the model to settle on the standard component ordering that it will encounter at inference. In the final stage, both component types and connections are masked simultaneously. The model now jointly decides which components to activate and how to connect them together.

For each input spec, AaLLM produces multiple candidate topologies, giving the downstream selection stage a diverse pool to choose from. The output of this stage is a ranked list of $K$ candidate topologies. The candidate $i$ is represented by a tuple $(\mathbf{B}_i, \boldsymbol{\tau}_i, r_i)$, where $\mathbf{B}_i$ is the bipartite connection matrix, $\boldsymbol{\tau}_i$ is the component-type vector, and $r_i$ is the preliminary SPICE simulation result obtained by instantiating the candidate with default parameters. To rank these candidates a weighted distance between their simulated results and the target spec is computed:
\begin{equation}\label{eq:candidate_score}
    \text{score}(\mathbf{B}_i) = \sum_{k} w_k \cdot
    \frac{|r_{i,k} - s^{*}_k|}{\max(|s^{*}_k| + \epsilon)}.
\end{equation}
Here, $r_{i,k}$ is the simulated value of spec $k$ for candidate $i$. Also, $w_k$ is a user-defined weight that reflects the relative importance of each spec. The denominator normalizes the deviation relative to the target, and $\epsilon$ > 0 is a small positive constant that is introduced to avoid instability. The resulting ranked list is not the final selection, which is made by the RAG module, but it provides a pre-filtered shortlist grounded in simulation-validated evidence. 

\subsection{RAG-Augmented Topology Selection}
\label{sec:rag_module}
The RAG module injects reasoning by letting an LLM selection agent consult an analog-design knowledge base before selecting a topology. The RAG module issues a query combining the candidate topologies, their preliminary results, and the user's original specs. This query is then searched against two separate indexes of the same knowledge base: one that matches by semantics and one that matches by exact keywords.

The semantic index stores contextually-augmented text chunks from research papers, textbooks, and PDK documents. These chunks are embedded as dense vectors for similarity search. The keyword index stores the same chunks under BM25~\cite{robertson2009probabilistic}, a classical ranking function that scores documents based on exact term matches. Running both semantic and keyword searches in parallel addresses a well-known weakness of pure semantic retrieval in technical domains. Semantic search often confuses related but distinct acronyms such ``CMRR'' and ``PSRR''. Keyword search handles exact terms reliably but misses conceptual paraphrases. The results from both indices are merged using a weighted reciprocal-rank fusion,
\begin{equation}\label{eq:rrf}
    \text{score}(d) = \alpha \cdot \frac{1}{\text{rank}_{\text{sem}}(d)+1}
    + (1-\alpha) \cdot \frac{1}{\text{rank}_{\text{bm25}}(d)+1},
\end{equation}
where $d$ denotes a document chunk in the retrieval set, $\text{rank}_{\text{sem}}(d)$ and $\text{rank}_{\text{bm25}}(d)$ are the positions of $d$ in the semantic and keyword search results respectively (with rank $1$ being the most relevant), and $\alpha \in [0, 1]$ is a weighting parameter that balances the contribution of each search method. The $+1$ term in each denominator prevents division by zero for the top-ranked document and reduces the influence of lower-ranked results.

A distinctive feature of our knowledge base construction is \emph{contextual augmentation}. Before embedding, each raw text chunk is passed through an LLM that creates a short description situating the chunk within its corresponding document. The tool set also includes a web-scraping action, which allows the agent to retrieve full papers from the open literature during its Reason+Act (ReAct) loop. It enables the agent to incorporate online literature search alongside the local knowledge base when needed.

\begin{table}[t]
\centering
\scriptsize
\setlength{\tabcolsep}{3pt}
\caption{AaLLM's models and settings per pipeline stage.}
\label{tab:models_settings}
\newcolumntype{L}[1]{>{\raggedright\arraybackslash\hspace{0pt}}p{#1}}
\begin{tabular}{@{}L{0.19\columnwidth}L{0.26\columnwidth}L{0.47\columnwidth}@{}}
\toprule
\textbf{Stage} & \textbf{Model} & \textbf{Decoding / settings} \\
\midrule
Topology generation &
FLAN-T5-base &
Constrained decoding; $T{=}0.8$, top-$k$ 50, top-$p$ 0.95, 256 tokens \\
\midrule
Topology selection (RAG) &
Claude Haiku 4.5 &
$T{=}0$, 1000 tokens \\
\midrule
Tri-agent sizing &
Claude Sonnet 4.6 &
$T{=}0.7$; 1024 tokens; $\leq$20 iterations \\
\midrule
RAG sizing knowledge &
Claude Sonnet 4.6 &
$T{=}0$, 4096 tokens \\
\midrule
RAG embed / rerank &
\texttt{voyage-2}, \texttt{rerank-\allowbreak english-\allowbreak v3.0} &
Corpus: Razavi textbook $+$ empirical sweep \\
\bottomrule
\end{tabular}
\end{table}

\begin{figure}[t]
\centering
\includegraphics[width=\columnwidth]{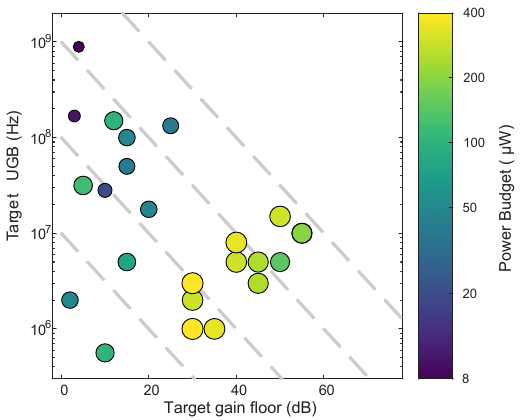}
\caption{Coverage of the 24 OPAMP target specs. Each circle represents a target gain and UGB; color and size encode the power budget. Dashed lines are constant gain-bandwidth-product contours.}
\label{fig:task_suite}
\end{figure}

\subsection{Tri-Agent Sizing Loop}\label{sec:tri_agent}
AaLLM employs three specialized LLM agents (Evaluator, Critic and Designer) operating in a closed loop around a SPICE simulator. Each agent plays a role that a human designer would apply in sequence: diagnose the current result (Critic), decide on a strategy (Evaluator), and execute a circuit parameter change (Designer). These roles are deliberately separated rather than collapsing them into a single agent, as each benefits from a different reasoning mode.

The \emph{Critic}, ``the eyes'', receives the current SPICE result and the target spec, and produces a structured diagnosis. In case the specs are not met, it reports a severity status (``on-target'', ``minor'', ``moderate'', or ``severe''), a relevant explanation, and the subset of components most directly responsible for the failure. 

The \emph{Evaluator}, ``the brain'', is one level above the loop and monitors the trajectory over recent iterations. It catches patterns that no single diagnosis can detect, such as a spec that stagnates or a trend that degrades over time. The Evaluator uses a two-tier design to keep costs low: a pre-check algorithm that scans the history at every iteration, and the Evaluator's LLM is invoked only when a pattern is detected or on a periodic audit. When the Evaluator is triggered, it instructs the Designer either to follow a specific action, temporarily ban certain circuit parameters from being modified, or override the current optimization strategy.

The \emph{Designer}, ``the hands'', is the only agent that actually touches parameters. It receives several inputs at each iteration: the Critic's diagnosis, any mandatory instructions and bans from the Evaluator, the current parameter values, and a short per-parameter trend history. Based on these inputs, the Designer emits a small list of modifications. Each modification specifies a component, a parameter, an action (\texttt{SET}, \texttt{SCALE}, or \texttt{OFFSET}), and a value. 

\subsection{Curriculum-Based Optimization}\label{sec:curriculum}
Analog circuit specs are correlated. In other words, changing one spec leads to a change in others. The curriculum controller enforces a process flow that mimics how a human designer would approach the problem. Optimization proceeds through three sequential phases: DC biasing, AC analysis and finally transient analysis. Each phase is promoted only when its objectives are met, and later phases run background checks that can demote the loop if earlier objectives regress beyond relaxed tolerances. The active curriculum phase is injected directly into the Critic's and Designer's prompts at every iteration, so that the agents reason about a locally simpler problem than the full multi-objective specs throughout the trajectory.

\begin{table}[t]
\centering
\caption{Qualitative comparison of AaLLM against SOTA approaches.}
\label{tab:SOTA_comparison_single}
\footnotesize
\setlength{\tabcolsep}{3pt}
\begin{tabular}{M{15mm}M{26mm}M{13mm}M{10mm}M{13mm}}
\textbf{Method} & \textbf{Topology Generation \& Sizing} & \textbf{Optimizer Free} & \textbf{RAG} & \textbf{Curriculum Sizing} \\
\midrule
Atelier~\cite{shen2025atelier} & \textnormal{\cmark} & \xmark & \cmark & \xmark\\
LaMAGIC~\cite{chang2024lamagic} & \xmark & \cmark & \xmark & \xmark\\
AmpAgent~\cite{liu2024ampagent}  & \xmark & \xmark & \cmark & \xmark\\
AnalogCoder~\cite{lai2025analogcoder} & \xmark & \cmark & \cmark & \xmark\\
\textbf{AaLLM} & \cmark & \cmark & \cmark & \cmark \\
\bottomrule
\end{tabular}
\end{table}

\section{Evaluation}
\label{sec:eval}
In this section, multiple operational amplifiers (OPAMPs) and filters are generated by AaLLM to evaluate its performance. The results are validated by SPICE simulations. The design parameters include transistor sizes ($W$, $L$, $V_{\text{bias}}$) and capacitor values. The general OPAMP desired target optimization problem are as follows: 
\begin{equation}
\begin{aligned}
\text{maximize} \quad & \{G,\, UGBW,\, PM\} \\
\text{minimize} \quad & \{I_{\text{bias}}\} \\
\text{subject to} \quad & G \geq G^{*},\; UGBW \geq UGBW^{*}, \\
& PM \geq PM^{*},\; I_{\text{bias}} \leq I_{\text{bias}}^{*}.
\end{aligned}
\label{eq:eval_optimization}
\end{equation}
Here, $(\cdot)$ and $(\cdot)^*$ denote the achieved spec and corresponding target spec, respectively. A solution is considered successful only if it meets all four target specs at the same time \textit{i.e.} $G~\geq~G^*$, $UGBW~\geq~UGBW^*$, $PM~\geq~PM^*$, and $I_{bias}~\leq~I_{bias}^*$. The circuits are synthesized using the 130nm SkyWater technology node~\cite{sky130_efabless}. Additionally, AaLLM employs different LLMs at each stage of the pipeline, selected according to that stage's requirements; these models are listed in Table~\ref{tab:models_settings}.

\begin{table}[t]
\centering
\caption{Ablation study on AaLLM conducted on a two-stage Miller OTA. PH: Physics heuristics; MC: Matching constraint.}
\label{tab:ablation}
\footnotesize
\setlength{\tabcolsep}{4pt}
\begin{tabular}{M{3mm}M{36mm}M{8mm}M{8mm}M{8mm}M{8mm}}
\toprule
\textbf{ID} & \textbf{Configuration} & \textbf{Iters} & \textbf{LLM calls} & \textbf{Gain (dB)} & \textbf{UGB (MHz)} \\
\midrule
D0 & Tri-Agent w/o RAG               & 17.3 & 49.7 & 48.1 & 11.24 \\
D1 & Tri-Agent w/o PH & 15.3 & 43.7 & 42.7 & 10.14 \\
D2 & Tri-Agent w/o MC & 20.0 & 56.0 & 31.8 & \phantom{0}5.14 \\
\textbf{D3} & \textbf{AaLLM (Tri-Agent $+$ RAG)} & \textbf{10.3} & \textbf{31.0} & \textbf{50.7} & \textbf{19.45} \\
\bottomrule
\end{tabular}
\end{table}

\subsection{Topology Generation Evaluation}
To evaluate the quality of the topology generator independently of sizing, we report Pass@$K$~\cite{lai2025analogcoder}. Pass@$K$ reports the fraction of sampled topologies that pass all validity checks. We report Pass@$1$. This 24 target specs OPAMP benchmark suite is depicted in Fig.~\ref{fig:task_suite}. The topology-generation is evaluated with Pass@1 and computed over 50 samples per task (5 seeds $\times$ 10 candidates). Runs with different seeds are statistically independent of each other. Each task is evaluated for structural validity defined by a valid OPAMP bi-partite matrix. The electrical validity is defined as a successful SPICE convergence with positive gain~(dB). AaLLM achieves \textbf{Pass@$1$ = 100\%} for both structural and electrical validity on all 24 target specs.

\begin{figure}[t]
\centering
\includegraphics[width=\columnwidth]{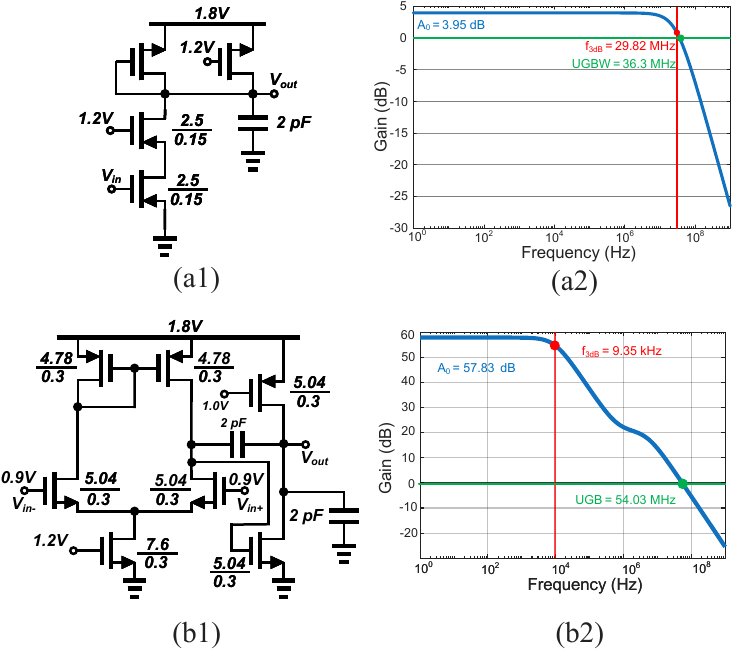}
\caption{Novel topologies generated and sized by AaLLM for a target design spec with transistor sizes shown as $\frac{W(\mu\text{m})}{L(\mu\text{m})}$. (a1) Single stage OPAMP schematic. (a2) Frequency response of the single stage OPAMP. (b1) Two stage OPAMP schematic. (b2) Frequency response of two stage OPAMP.
}
\label{fig:schematic_novel}
\end{figure}

\subsection{Novel Topology Generation}
A key advantage of AaLLM over library-based approaches is its ability to produce novel topologies that do not appear in the training corpus. Fig.~\ref{fig:schematic_novel} shows single stage and two-stage novel OPAMP topologies generated by AaLLM. 

\subsection{Circuit Sizing Evaluation}
Out of the 24 target specs depicted in Fig.~\ref{fig:task_suite}, 12 (gain 30--70~dB) are the hardest to achieve. To test AaLLM's circuit sizing, we evaluate these 12 tasks against both a novel two-stage topology (Fig.~\ref{fig:schematic_novel}) and a known canonical two-stage topology. Design quality is summarized by a figure of merit (FoM) that averages the normalized margin across the three targets,

\begin{table}[t]
\centering
\footnotesize
\renewcommand{\arraystretch}{1.25}
\setlength{\tabcolsep}{8pt}
\caption{Head-to-head AaLLM sizing on 12 tasks, two seeds each. Tested on a
two-stage novel topology generated by AaLLM and a canonical Miller OPAMP.
\\[4pt]}
\label{tab:novel_vs_miller}
\begin{tabular}{@{}c cc cc@{}}
\toprule
& \multicolumn{2}{c}{\textbf{Novel}} & \multicolumn{2}{c}{\textbf{Canonical Miller}} \\
\cmidrule(lr){2-3}\cmidrule(l){4-5}
\textbf{Task} & \textbf{FoM} & \textbf{Met} & \textbf{FoM} & \textbf{Met} \\
\midrule
 1 & 0.550 & 2/2 & 0.791 & 2/2 \\
 2 & 0.428 & 2/2 & 0.628 & 2/2 \\
 3 & 0.583 & 2/2 & 0.452 & 2/2 \\
 4 & 0.461 & 2/2 & 0.618 & 2/2 \\
 5 & 0.442 & 2/2 & 0.528 & 2/2 \\
 6 & 0.368 & 2/2 & 0.453 & 2/2 \\
 7 & 0.463 & 2/2 & 0.488 & 1/2 \\
 8 & 0.355 & 2/2 & 0.432 & 2/2 \\
 9 & 0.295 & 2/2 & 0.334 & 2/2 \\
10 & 0.185 & 0/2 & 0.296 & 2/2 \\
11 & 0.318 & 2/2 & 0.108 & 1/2 \\
12 & 0.281 & 2/2 & 0.324 & 2/2 \\
\midrule
\textbf{Mean / total} & \textbf{0.394} & \textbf{22/24} & \textbf{0.454} & \textbf{22/24} \\
\bottomrule
\end{tabular}
\end{table}

\begin{equation}
\mathrm{FoM} = \frac{1}{3}\!\left[
\frac{A_v - A_{v,\min}}{A_v} +
\frac{f_u - f_{u,\min}}{f_u} +
\frac{P_{\max} - P}{P_{\max}}\right],
\label{eq:fom}
\end{equation}
where $A_{v,\min}$, $f_{u,\min}$, and $P_{\max}$ are the specified gain, unity-gain bandwidth (UGB), and power budget. A positive term indicates the corresponding spec is met with margin, so a higher FoM denotes a design that clears all three targets more comfortably. The canonical Miller OTA meets \textbf{all 12/12 target specs} (best of two seeds), with 22 of 24 runs succeeding overall. As shown in Table~\ref{tab:novel_vs_miller}, both misses (Tasks 7 and 11) are seed variance: the alternate seed meets specs in each case, so the miss is recoverable. The novel topology attains a comparable FoM, also succeeding on 22 of 24 runs, though these resolve to 11 of 12 tasks; the lone unrecovered miss is the tightest gain--power corner.

\subsection{Comparison Against State of the Art}
Table~\ref{tab:SOTA_comparison_single} compares AaLLM qualitatively against four recent LLM-based analog design frameworks. Atelier couples a multi-agent flow to a curated knowledge base, but delegates sizing to CMA-ES~\cite{shen2025atelier}. AmpAgent retrieves design equations from literature, yet relies on manual netlist entry and conventional optimizers such as ABC and TuRBO~\cite{liu2024ampagent}. LaMagic~\cite{chang2024lamagic} generates topologies in a single pass, but is trained by supervised fine-tuning and never sizes device parameters. AnalogCoder~\cite{lai2025analogcoder} reuses sub-circuits from a tool library while explicitly avoiding parameter optimization. AaLLM is the only framework that closes the loop end to end: it generates topology, sizes devices without an external optimizer, and retrieves design knowledge through RAG. Curriculum-based sizing is unique to our approach.

\begin{table}[t]
\centering
\footnotesize
\caption{AaLLM vs.\ Atelier on four two-stage circuits, two
seeds each (per-circuit time budget 200--4000\,s).}
\label{tab:cmaes_compare}
\begin{tabular}{@{}ll cc cc@{}}
\toprule
& & \multicolumn{2}{c}{\textbf{AaLLM}} & \multicolumn{2}{c}{\textbf{Atelier~\cite{shen2025atelier}}} \\
\cmidrule(lr){3-4}\cmidrule(l){5-6}
\textbf{Task} & \textbf{Topology} & \textbf{Met} & \textbf{\#Simulations} & \textbf{Met} & \textbf{\#Simulations} \\
\midrule
1 & Miller & \textbf{2/2} & 350 & 1/2 & 1382 \\
2 & Miller & \textbf{2/2} & 294 & 1/2 & \phantom{0}722 \\
3 & Novel  & \textbf{2/2} & 154 & 1/2 & \phantom{0}122 \\
4 & Novel  & \textbf{2/2} & 112 & 1/2 & \phantom{0}314 \\
\midrule
\multicolumn{2}{@{}l}{\textbf{Total}} & \textbf{8/8} & 910 & \textbf{4/8} & 2540 \\
\bottomrule
\end{tabular}
\end{table}

\begin{table}[t]
\centering
\footnotesize
\setlength{\tabcolsep}{4pt}
\caption{Active filters: target vs.\ AaLLM-achieved specs for three topologies: Sallen-Key low-pass (SK\_LPF), Sallen-Key high-pass (SK\_HPF), and multi-feedback band-pass (MFB\_BPF).}
\label{tab:filter_extension}
\begin{tabular}{@{}c cc cc cc@{}}
\toprule
\textbf{Spec} & \multicolumn{2}{c}{\textbf{SK\_LPF}} & \multicolumn{2}{c}{\textbf{SK\_HPF}} & \multicolumn{2}{c}{\textbf{MFB\_BPF}} \\
\cmidrule(lr){2-3}\cmidrule(lr){4-5}\cmidrule(l){6-7}
& \textbf{Target} & \textbf{Result} & \textbf{Target} & \textbf{Result} & \textbf{Target} & \textbf{Result} \\
\midrule
Gain (dB)        & $\geq 0$  & 1.65 & $\geq 0$  & 1.92 & $\geq 18$ & 21.5 \\
Freq (kHz)       & $\geq 8$  & 8.08 & $\geq 4$  & 5.16 & $\geq 9$  & 9.96 \\
Selectivity (dB) & $\geq 18$ & 27.6 & $\geq 18$ & 31.6 & $\geq 4$  & 4.33 \\
Power ($\mu$W)   & $\leq 300$ & 293 & $\leq 300$ & 280 & $\leq 250$ & 123 \\
\bottomrule
\end{tabular}
\end{table}

\begin{figure}[t]
\centering
\includegraphics[width=\columnwidth]{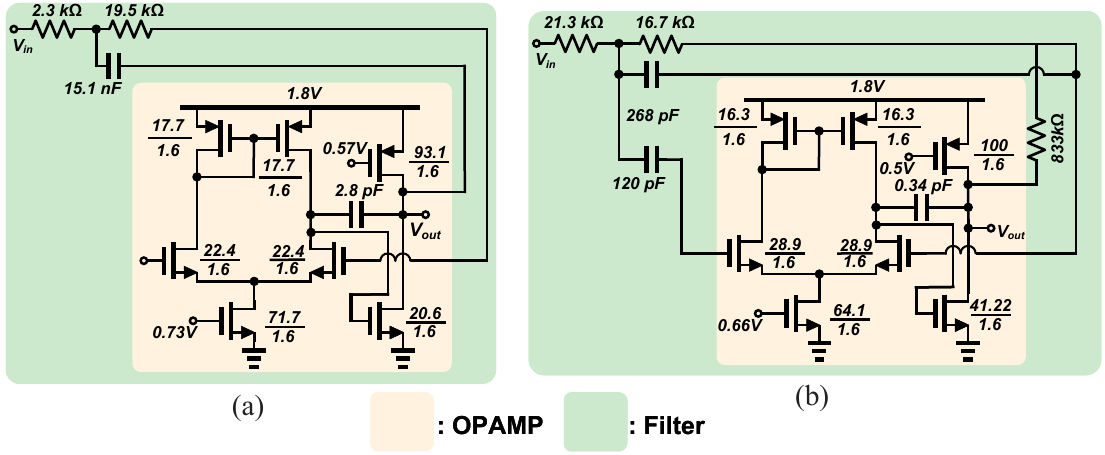}
\caption{Active-filter schematics sized by AaLLM.
(a)  Sallen-Key low-pass filter (b) Multi-feedback band-pass filter. 
}
\label{fig:filters}
\end{figure}

To test against SOTA, we ported AaLLM to the Arizona State University Predictive Technology Model (ASU PTM) 45\,nm PDK to enable a direct comparison against AnaFlow~\cite{ahmadzadeh2025anaflow}, a recent LLM-based sizing framework, on the same technology node and benchmark. On the direct AnaFlow benchmark, a two-stage canonical Miller OPAMP, AaLLM meets the target specs (gain, UGB, power) while calling SPICE twice. On the other hand, it takes AnaFlow 9 SPICE calls using its best-performing LLM backbone (Gemini 2.5 Pro). In other words, the results show a \textbf{4.5x} decrease in number of SPICE call and a \textbf{40x} decrease in wall-clock time compared to AnaFlow. To broaden our comparison beyond AnaFlow, we also benchmark against Atelier~\cite{shen2025atelier}, under wall-clock-matched conditions. The time budget for this test ranged between 200--4000s. The circuit sizing mechanism used in Atelier is conventional optimizer, the Covariance Matrix Adaptation Evolution Strategy (CMA-ES). As shown in Table~\ref{tab:cmaes_compare}, AaLLM meets the target specs on all 8 circuits where as Atelier meets only 4 of them. Moreover, AaLLM uses \textbf{2.79x} fewer circuit simulations.

\subsection{Ablation Study}\label{sec:ablation}
To isolate the contribution of each stage, we ablate several variants of AaLLM on a Miller OTA with target specs (gain $\geq 46$\,dB, UGB $\geq 5$\,MHz, power $\leq 150\,\mu\text{W}$). Each variant is run with a 20-iteration budget. As shown in Table~\ref{tab:ablation}, (D0) is the tri-agent loop without RAG; (D1) removes the physics heuristics from the Designer prompt; (D2) removes the differential-pair matching constraint used to ensure a valid structural topology. The full AaLLM configuration (D3) improves UGB by \textbf{73\%} while reducing the iteration count by \textbf{40\%}.

\subsection{AaLLM Scaled to Active Filters}
Beyond the OPAMP suite, we extended the same bipartite-matrix representation and tri-agent sizing flow to a second family; active filters across three topologies: Sallen-Key low-pass (SK\_LPF), Sallen-Key high-pass (SK\_HPF), and multi-feedback band-pass (MFB\_BPF), as illustrated in Fig.~\ref{fig:filters}. Each is sized to meet all target specs, jointly tuning transistor widths and passive resistors/capacitors. Achieved-versus-target specs of one sample for all three topologies are listed in Table~\ref{tab:filter_extension}. As shown, AaLLM meets all three target specs.

\section{Conclusion}
In this work, we introduce AaLLM, an open-source, fully-automated framework for analog circuit design that handles both topology generation and circuit sizing in a single automated pipeline. A fine-tuned FLAN-T5 model generates candidate topologies as bipartite component-node matrices, a RAG module selects the most suitable topology using established circuit theory, and a tri-agent sizing loop refines the circuit parameters through SPICE-in-the-loop simulation. AaLLM is capable of generating electrically functional novel topologies beyond its training corpus. Our results show a 3x - 4.5x decrease in SPICE simulations and a 40x decrease in wall-clock time using AaLLM compared to SOTA methods.

\bibliographystyle{IEEEtran}
%\bibliography{IEEEabrv,Bibliography}

\begin{thebibliography}{10}
\providecommand{\url}[1]{#1}
\csname url@samestyle\endcsname
\providecommand{\newblock}{\relax}
\providecommand{\bibinfo}[2]{#2}
\providecommand{\BIBentrySTDinterwordspacing}{\spaceskip=0pt\relax}
\providecommand{\BIBentryALTinterwordstretchfactor}{4}
\providecommand{\BIBentryALTinterwordspacing}{\spaceskip=\fontdimen2\font plus
\BIBentryALTinterwordstretchfactor\fontdimen3\font minus \fontdimen4\font\relax}
\providecommand{\BIBforeignlanguage}[2]{{%
\expandafter\ifx\csname l@#1\endcsname\relax
\typeout{** WARNING: IEEEtran.bst: No hyphenation pattern has been}%
\typeout{** loaded for the language `#1'. Using the pattern for}%
\typeout{** the default language instead.}%
\else
\language=\csname l@#1\endcsname
\fi
#2}}
\providecommand{\BIBdecl}{\relax}
\BIBdecl

\bibitem{fayazi_2021}
M.~Fayazi, Z.~Colter, E.~Afshari, and R.~Dreslinski, ``Applications of Artificial Intelligence on the Modeling and Optimization for Analog and Mixed-Signal Circuits: A Review,'' \emph{IEEE Transactions on Circuits and Systems I: Regular Papers}, vol.~68, no.~6, pp. 2418--2431, 2021.

\bibitem{Ren_2020}
H.~Ren, G.~F. Kokai, W.~J. Turner, and T.-S. Ku, ``ParaGraph: Layout Parasitics and Device Parameter Prediction using Graph Neural Networks,'' in \emph{Proceedings of the 57th ACM/EDAC/IEEE Design Automation Conference}, ser. DAC '20.\hskip 1em plus 0.5em minus 0.4em\relax IEEE Press, 2020.

\bibitem{Settaluri_2020}
K.~Settaluri, A.~Haj~Ali, Q.~Huang, K.~Hakhamaneshi, and B.~Nikolic, ``AutoCkt: Deep Reinforcement Learning of Analog Circuit Designs,'' 01 2020.

\bibitem{brempong2026oracle}
O.~Brempong, M.~A. Habib, V.~Poddar, and M.~Fayazi, ``ORACLE: A Multi-Objective Reinforcement Learning-Based Analog Circuit Design Optimizer with Large Language Models-Guided Exploration,'' \emph{arXiv preprint arXiv:2608.04999}, 2026.

\bibitem{fayazi2022fascinet}
M.~Fayazi \emph{et~al.}, ``FASCINET: A Fully Automated Single-Board Computer Generator Using Neural Networks,'' \emph{IEEE Transactions on Computer-Aided Design of Integrated Circuits and Systems (TCAD)}, 2022.

\bibitem{Lyu_2018}
W.~Lyu, P.~Xue, F.~Yang, C.~Yan, Z.~Hong, X.~Zeng, and D.~Zhou, ``An Efficient Bayesian Optimization Approach for Automated Optimization of Analog Circuits,'' \emph{IEEE Transactions on Circuits and Systems I: Regular Papers}, vol.~65, no.~6, pp. 1954--1967, 2018.

\bibitem{fayazi2023angel}
M.~Fayazi, M.~T. Taba, E.~Afshari, and R.~Dreslinski, ``AnGeL: Fully-Automated Analog Circuit Generator Using a Neural Network Assisted Semi-Supervised Learning Approach,'' \emph{IEEE Transactions on Circuits and Systems I: Regular Papers}, vol.~70, no.~11, pp. 4516--4529, 2023.

\bibitem{abbineni2026muallm}
P.~Abbineni~\textit{et al.}, ``MuaLLM: A Multimodal Large Language Model Agent for Circuit Design Assistance with Hybrid Contextual Retrieval-Augmented Generation,'' in \emph{2026 31st Asia and South Pacific Design Automation Conference (ASP-DAC)}.\hskip 1em plus 0.5em minus 0.4em\relax IEEE, 2026, pp. 646--652.

\bibitem{aldowaish2026sina}
S.~Aldowaish, Y.~Karumanchi, K.-C. Chiang, S.~Noorzad, and M.~Fayazi, ``SINA: A Circuit Schematic Image-to-Netlist Generator Using Artificial Intelligence,'' \emph{Design, Automation and Test in Europe Conference and Exhibition (DATE)}, 2026.

\bibitem{lewis2020rag}
P.~Lewis, E.~Perez, A.~Piktus, F.~Petroni, V.~Karpukhin, N.~Goyal, H.~K{\"u}ttler, M.~Lewis, W.-t. Yih, T.~Rockt{\"a}schel, S.~Riedel, and D.~Kiela, ``Retrieval-Augmented Generation for Knowledge-Intensive NLP Tasks,'' in \emph{Advances in Neural Information Processing Systems (NeurIPS)}, vol.~33, 2020, pp. 9459--9474.

\bibitem{devlin2019bert}
J.~Devlin, M.-W. Chang, K.~Lee, and K.~Toutanova, ``BERT: Pre-training of Deep Bidirectional Transformers for Language Understanding,'' in \emph{Proceedings of the 2019 conference of the North American chapter of the association for computational linguistics: human language technologies, volume 1 (long and short papers)}, 2019, pp. 4171--4186.

\bibitem{vaswani2017attention}
A.~Vaswani, N.~Shazeer, N.~Parmar, J.~Uszkoreit, L.~Jones, A.~N. Gomez, {\L}.~Kaiser, and I.~Polosukhin, ``Attention is All you Need,'' in \emph{Advances in Neural Information Processing Systems (NeurIPS)}, vol.~30, 2017, pp. 5998--6008.

\bibitem{chen2025analogseeker}
Z.~Chen~\textit{et al.}, ``AnalogSeeker: An Open-source Foundation Language Model for Analog Circuit Design,'' \emph{arXiv preprint arXiv:2508.10409}, 2025.

\bibitem{chang2024lamagic}
C.-C. Chang~\textit{et al.}, ``LaMAGIC: Language-Model-based Topology Generation for Analog Integrated Circuits,'' \emph{arXiv preprint arXiv:2407.18269}, 2024.

\bibitem{shi2025amsnet}
Y.~Shi~\textit{et al.}, ``AMSnet-KG: A Netlist Dataset for LLM-based AMS Circuit Auto-design Using Knowledge Graph RAG,'' \emph{ACM Transactions on Design Automation of Electronic Systems}, vol.~30, no.~6, pp. 1--37, 2025.

\bibitem{liu2024ampagent}
C.~Liu~\textit{et al.}, ``AmpAgent: An LLM-based Multi-Agent System for Multi-stage Amplifier Schematic Design from Literature for Process and Performance Porting,'' \emph{arXiv preprint arXiv:2409.14739}, 2024.

\bibitem{lai2025analogcoder}
Y.~Lai~\textit{et al.}, ``AnalogCoder: Analog Circuit Design via Training-Free Code Generation,'' in \emph{Proceedings of the AAAI Conference on Artificial Intelligence}, vol.~39, no.~1, 2025, pp. 379--387.

\bibitem{gao2025analoggenie}
J.~Gao, W.~Cao, J.~Yang, and X.~Zhang, ``AnalogGenie: A Generative Engine for Automatic Discovery of Analog Circuit Topologies,'' \emph{arXiv preprint arXiv:2503.00205}, 2025.

\bibitem{zhang2025analogxpert}
H.~Zhang, S.~Sun, Y.~Lin, R.~Wang, and J.~Bian, ``AnalogXpert: Automating Analog Topology Synthesis by Incorporating Circuit Design Expertise into Large Language Models,'' in \emph{2025 International Symposium of Electronics Design Automation (ISEDA)}.\hskip 1em plus 0.5em minus 0.4em\relax IEEE, 2025, pp. 772--777.

\bibitem{vijayaraghavan2025autocircuit}
P.~Vijayaraghavan, L.~Shi, E.~Degan, V.~Mukherjee, and X.~Zhang, ``AUTOCIRCUIT-RL: Reinforcement Learning-Driven LLM for Automated Circuit Topology Generation,'' \emph{arXiv preprint arXiv:2506.03122}, 2025.

\bibitem{ahmadzadeh2025anaflow}
M.~Ahmadzadeh, K.~Chen, and G.~Gielen, ``AnaFlow: Agentic LLM-based Workflow for Reasoning-Driven Explainable and Sample-Efficient Analog Circuit Sizing,'' in \emph{2025 IEEE/ACM International Conference On Computer Aided Design (ICCAD)}.\hskip 1em plus 0.5em minus 0.4em\relax IEEE, 2025, pp. 1--7.

\bibitem{yu2026autosizer}
X.~Yu, D.~Torbunov, S.~Mandal, and Y.~Ren, ``AutoSizer: Automatic Sizing of Analog and Mixed-Signal Circuits via Large Language Model (LLM) Agents,'' \emph{arXiv preprint arXiv:2602.02849}, 2026.

\bibitem{wu2025easysize}
X.~Wu, F.~Hu, S.~J. Babu, Y.~Zhao, and X.~Guo, ``EasySize: Elastic Analog Circuit Sizing via LLM-Guided Heuristic Search,'' \emph{arXiv preprint arXiv:2508.05113}, 2025.

\bibitem{kochar2025ledro}
D.~V. Kochar, H.~Wang, A.~P. Chandrakasan, and X.~Zhang, ``LEDRO: LLM-Enhanced Design Space Reduction and Optimization for Analog Circuits,'' in \emph{2025 IEEE International Conference on LLM-Aided Design (ICLAD)}.\hskip 1em plus 0.5em minus 0.4em\relax IEEE, 2025, pp. 141--148.

\bibitem{ghosh2025accelerating}
S.~Ghosh, E.~Y. Gebru, C.~V. Kashyap, R.~Harjani, and S.~S. Sapatnekar, ``Accelerating OTA Circuit Design: Transistor Sizing Based on a Transformer Model and Precomputed Lookup Tables,'' in \emph{2025 Design, Automation \& Test in Europe Conference (DATE)}.\hskip 1em plus 0.5em minus 0.4em\relax IEEE, 2025, pp. 1--7.

\bibitem{lai2026analogcoder}
Y.~Lai~\textit{et al.}, ``AnalogCoder-Pro: Unifying Analog Circuit Generation and Optimization via Multi-modal LLMs,'' \emph{IEEE Transactions on Computer-Aided Design of Integrated Circuits and Systems}, 2026.

\bibitem{chen2024artisan}
Z.~Chen, J.~Huang, Y.~Liu, F.~Yang, L.~Shang, D.~Zhou, and X.~Zeng, ``Artisan: Automated Operational Amplifier Design via Domain-specific Large Language Model,'' in \emph{Proceedings of the 61st ACM/IEEE Design Automation Conference}, 2024, pp. 1--6.

\bibitem{shen2025atelier}
J.~Shen~\textit{et al.}, ``Atelier: An Automated Analog Circuit Design Framework via Multiple Large Language Model-Based Agents,'' \emph{IEEE Transactions on Computer-Aided Design of Integrated Circuits and Systems}, 2025.

\bibitem{liu2024ladac}
C.~Liu, Y.~Liu, Y.~Du, and L.~Du, ``LADAC: Large Language Model-driven Auto-Designer for Analog Circuits,'' \emph{Authorea Preprints}, 2024.

\bibitem{chen2025menter}
P.-H. Chen~\textit{et al.}, ``MenTeR: A fully-automated Multi-agenT workflow for end-to-end RF/Analog Circuits Netlist Design,'' in \emph{2025 IEEE International Conference on LLM-Aided Design (ICLAD)}.\hskip 1em plus 0.5em minus 0.4em\relax IEEE, 2025, pp. 124--132.

\bibitem{dong2023cktgnn}
Z.~Dong~\textit{et al.}, ``CktGNN: Circuit Graph Neural Network for Electronic Design Automation,'' \emph{arXiv preprint arXiv:2308.16406}, 2023.

\bibitem{wolf2020transformers}
T.~Wolf~\textit{et al.}, ``Transformers: State-of-the-Art Natural Language Processing,'' in \emph{Proceedings of the 2020 conference on empirical methods in natural language processing: system demonstrations}, 2020, pp. 38--45.

\bibitem{robertson2009probabilistic}
S.~Robertson and H.~Zaragoza, \emph{
The Probabilistic Relevance Framework BM25 and Beyond}.\hskip 1em plus 0.5em minus 0.4em\relax Now Publishers Inc, 2009, vol.~4.

\bibitem{sky130_efabless}
{SkyWater Technology Foundry}, {Google LLC}, and {Efabless Corporation}, ``{SkyWater Open Source PDK 130nm Process Design Kit},'' \url{https://skywater-pdk.readthedocs.io/}, 2022, accessed: \today.

\end{thebibliography}

\end{document}